\documentclass[preprint,aps,prl]{revtex4}
\usepackage{epsfig}
\usepackage{float}

\begin{document}

\title{Van der Waals interactions in DFT made easy by Wannier functions} 
\author{Pier Luigi Silvestrelli}

\affiliation{
Dipartimento di Fisica ``G. Galilei'',
Universit\`a di Padova, via Marzolo 8, I-35131 Padova, Italy
and DEMOCRITOS National Simulation Center, Trieste, Italy}

\date{\today}

\begin{abstract}
Ubiquitous Van der Waals interactions between atoms and molecules
are important for many molecular and solid structures.
These systems are often studied from first principles using
the Density Functional Theory (DFT). However, the commonly used
DFT functionals fail to capture the essence of
Van der Waals effects. Many attempts to correct for this
problem have been proposed, which are not completely
satisfactory because they are either very complex and computationally
expensive or have a basic semiempirical character.
We here describe a novel approach, based on the use
of the Maximally-Localized Wannier functions, that appears
to be promising, being simple, efficient, accurate, and
transferable (charge polarization effects are naturally included).
The results of test applications are presented.
\end{abstract}
\maketitle

\vfill \eject

\noindent
\narrowtext

DFT represents a well-established tool to study 
the structural and electronic properties of molecules and
condensed matter systems from first principles, and to 
elucidate complex processes such
as surface adsorptions, catalytic reactions, and diffusive motions. 
Although current density functionals are able to describe well
several systems, at much lower computational cost compared to other 
first principles methods, they fail to do so\cite{Kohn} for the 
description of long-range dispersion effects, generally
denoted as Van der Waals (VdW) interactions, particularly the leading 
$R^{-6}$ term originated from correlated instantaneous dipole 
fluctuations; the cases where DFT (using, for instance the 
Local Density Approximation, LDA, or the PBE\cite{PBE} functionals) 
provides reasonable estimates for the interaction
energy of weakly bound system are actually due to favorable errors
or cancellations and should therefore be considered accidental.

In order to overcome this severe deficiency of DFT, two basic
strategies have been adopted: on one hand, new density functionals
or/and relatively complex schemes have been proposed that in principle 
allow for a 
correct treatment of the VdW interactions
\cite{Kohn,Ashcroft,Langreth96,Hult,Dobson,Gonze,Langreth04,Grimme07},
on the other hand several semiempirical approaches\cite{Scoles,Wu}
have been developed where an approximately derived $R^{-6}$ term,
multiplied by a suitable short-range damping function,  
is explicitly introduced. Although both these approaches have been
somehow successful, neither of them appears to be entirely satisfactory:
in fact, the former is generally quite complex and computationally
very demanding, compared to a standard DFT calculation, while the latter,
based on interatomic $C_6$ coefficients (actually dependent on the 
molecular environment of the atoms involved) and empirical fits,
turns out to be far from generally applicable because it 
neglects changes in the atomic polarizabilities (which, in general, are
not additive) and should be 
tailored to the specific system considered. 
Therefore the development of a practical efficient scheme to include 
VdW interactions in DFT remains a great theoretical challenge.

In this paper we propose a novel method which allows the 
efficient calculation of the VdW interaction between nonoverlapping
fragments, using as input only the ground state electron 
density and the Kohn-Sham (KS) orbitals computed in a conventional
DFT approach.

Crucial to our analysis is the use of the Maximally-Localized 
Wannier function (MLWF) formalism\cite{Wannier}, that allows
the total electronic density to be partitioned, in a chemically 
transparent and unambiguous way, into individual fragment contributions.
The MLWFs represent a generalization, for
systems characterized by periodic boundary conditions, of the 
Boys' localized orbitals\cite{Boys} that are commonly used in quantum
chemistry; they allow for an intuitive interpretation of the 
bonding properties of condensed-matter systems\cite{Wannier} 
and are at the center of the modern theory of polarization\cite{Vanderbilt}.
The MLWFs, $\{w_n({\bf r})\}$, are generated by performing a unitary 
transformation in the subspace of the occupied KS orbitals, 
obtained by a standard DFT calculation, so as to minimize the total spread:

\begin{equation}
S = \sum_n S_n = 
\sum_n \left( \left<w_n|r^2|w_n\right> - 
\left<w_n|{\bf r}|w_n\right>^2 \right)\;.
\label{spread}
\end{equation}
Besides its spread, $S_n$, each MLWF is characterized also by its
Wannier-function center (WFC); for instance, 
if periodic boundary conditions are used with a cubic supercell of side $L$, 
the coordinate $x_n$ of the $n$-th WFC is defined\cite{Wannier} as

\begin{equation}
x_n = -{L\over {2\pi}}{\rm Im}\; {\rm ln}
\left< w_n | e^{-i{2\pi \over L} x} | w_n \right>\;,
\label{rcenter}
\end{equation}
with similar definitions for $y_n$ and $z_n$.
If spin degeneracy is exploited, every MLWF corresponds
to 2 paired electrons.
Starting from these MLWFs the leading
$R^{-6}$ VdW correction term can be evaluated using different possible
recipes; one of them is described and applied in the following. 
We make the reasonable (at least for insulating systems) 
assumption\cite{Resta} of exponential localization of the MLWFs
in real space, so that each of them is supposed to be an hydrogen-like,
normalized, function, centered around its WFC position, $r_n$, 
with a spread $S_n$:

\begin{equation}
w_n(|{\bf r - \bf r_n}|) = {3^{3/4} \over {\sqrt{\pi}S_n^{3/2}}}
e^{-{\sqrt{3}\over S_n}|{\bf r - \bf r_n}|}\;.
\label{MLWF}
\end{equation}
Then the binding energy of a system composed of two fragments is given by
$E_b=E_0+E_{\rm VdW}$, where $E_0$ is the binding energy obtained 
from a standard DFT calculation, while the VdW correction is 
assumed to have the form:

\begin{equation}
E_{\rm VdW} = -\sum_{n,l} f_{nl}(r_{nl}){C_{6nl}\over {r_{nl}^6}}\;,
\label{EVdw}
\end{equation}
where $r_{nl}$ is the 
distance of the $n$-th WFC, of the first fragment, 
from the $l$-th WFC of the second one, the sum is over all the 
MLFWs of the two fragments, and the $C_{6nl}$ 
coefficients can be calculated directly 
from the basic information (center positions and spreads) given
by the MLFWs. In fact, using for instance the expression proposed by 
Andersson {\it et al.} (see Eq. (10) of ref.\cite{Langreth96})
that describes the long-range interaction between two
separated fragments of matter:

\begin{equation}
C_{6nl}={3 \over {32\pi^{3/2}}}\int_{|{\bf r}|\leq r_c}d{\bf r}
\int_{|{\bf r'}|\leq r_c'}d{\bf r'}
{\sqrt{\rho_n(r)\rho_l(r')} \over {\sqrt{\rho_n(r)}+\sqrt{\rho_l(r')}}}
={3 \over {32\pi^{3/2}}}\int_{|{\bf r}|\leq r_c}d{\bf r}
\int_{|{\bf r'}|\leq r_c'}d{\bf r'}
{{w_n(r)w_l(r')} \over {w_n(r)+w_l(r')}}\;,
\label{integral}
\end{equation}
where $\rho_n(r) = w_n^2(r)$ is the electronic density corresponding
to the $n$-th MLWF, $C_{6nl}$ is given in a.u., and
the $r_c$, $r_c'$ cutoffs have been 
introduced\cite{Ashcroft,Langreth96}
to properly take into account both the limit of
separated fragments and of distant disturbances in an electron gas:
by equating the length scale for density change to the
electron gas screening length one obtains:

\begin{equation}
{{6\rho_n(r_c)}\over {|\vec{\bigtriangledown}\rho_n(r_c)|}} =
{{v_F[\rho_n(r_c)]} \over {\omega_p[\rho_n(r_c)]}}\;,
\label{cutoff1}
\end{equation}
where $v_F = \left(3\pi^2\rho_n(r) \right)^{1/3}/m$
is the local Fermi velocity, and $\omega_p = 
\left( {{4\pi e^2 \rho_n(r)}/m} \right)^{1/2}$ is the 
local plasma frequency.
By using the analytic form (see Eq. (\ref{MLWF})) of the MLWFs,
it is straightforward to obtain the 
cutoff expressed in terms of the MLWF spread:

\begin{equation}
r_c = S_n\sqrt{3}\left(0.769 + 1/2{\rm ln}(S_n) \right)\;,
\label{cutoff2}
\end{equation}
and to evaluate very efficiently the multimensional integral
of Eq. (\ref{integral}). 
For instance, in the test case of 2, distant H atoms, using the well known 
(unperturbed) analytic H atom wavefunction, the above formula
gives $C_6 = 6.41$ a.u. to be compared to the reference literature value
of 6.50 a.u.

In Eq. (\ref{integral}), if the electronic density corresponding to 
every MLWF is multipled by 2, the $C_{6nl}$ coefficients increase
by a $\sqrt{2}$ factor; therefore  
it appears reasonable to assume that, when each
MLWF describes 2 paired electrons (spin degeneracy),
$C_{6nl}$ has to be multipled by $\sqrt{2}$.  
This is also supported by the fact that, in the Slater-Kirkwood 
approximation for estimating the $C_6$ coefficients, the
effective number of electrons is
smaller than the number of valence electrons, and it is
1.42 $\simeq \sqrt{2}$ in the case of the He atom\cite{Halgren},
whose DFT ground state is just given by 2 paired electrons 
in the lowest-energy KS orbital.

In Eq. (\ref{EVdw}) $f_{nl}(r)$ is a damping function which serves to cutoff
the unreasonable behavior of the asymptotic VdW correction at small 
fragment separations.
For it we have chosen a form\cite{Wu,Grimme} with parameters 
directly related to the MLWF spreads:

\begin{equation}
f_{nl}(r) = {1 \over {1+exp(-a(r/R_s-1))}}\;,
\label{fr}
\end{equation}
where\cite{Grimme} $a \simeq 20$ (the results are almost
independent on the particular value of this parameter), 
and $R_s = R_{\rm VdW}+R'_{\rm VdW}$ is the sum of the VdW radii 
of the MLWFs, which, following Grimme {\it et al.}\cite{Grimme},
are determined as the radii of the 0.01 electron density contour;
using Eq. (\ref{MLWF}) one easily obtains that:

\begin{equation}
R_{\rm VdW} = (1.475-0.866{\rm ln}(S_n))S_n\;.
\label{RVdw}
\end{equation}  
The damping function effectively reduces the VdW correction to
zero typically below 2 \AA; at intermediate distances a minimum
in the VdW potential exists that usually lies slightly below the
sum of the corresponding VdW radii, $R_s$.
Note that the above recipe resembles that proposed in ref.
\cite{Angyan}, where the long-range electron-electron interaction 
is separated by the short-range one, using a single parameter
describing the physical dimensions of a valence electron pair.

The $E_0$ binding energy can be obtained 
from a standard DFT calculation (we have used the
CPMD\cite{CPMD} and $\nu$-ESPRESSO\cite{ESPRESSO} ab initio packages),
using the Generalized Gradient Approximation (GGA) in the
revPBE flavor\cite{revPBE}. This choice\cite{Langreth04,Grimme} is motivated 
by the fact that revPBE is fitted to the exact Hartree-Fock exchange,
so that the VdW binding, a correlation effect, only comes
from the VdW correction term, as described above, without any
double-counting effect (for instance, LDA, or other GGA functionals, such
as PBE, predict substantial binding in rare gas dimers, due
to a severe overestimate of the long-range part of the 
exchange contribution\cite{Langreth04}).
The evaluation of the VdW correction as a post-standard DFT
perturbation, using the revPBE electronic density distribution,
represents an approximation because, in principle a full
self-consistent calculations should be performed; however
recent investigations\cite{Langreth07} on different systems 
have shown that the effects due to the lack of self-consistency
are negligible
(this is reasonable because one does not expect that the rather
weak and diffuse VdW interaction substantially changes the
electronic charge distribution).

The VdW correction scheme described above can be refined by
considering the effects due to the anisotropy of the MLWFs,
and distinguishing between contributions along (or orthogonal
to) the fragment-fragment direction (details will be published
elsewhere\cite{ELSE}). 
Moreover, also higher-order term VdW corrections, involving
the $C_8$, $C_{10}$,... coefficients, could be easily included.
Clearly, in the present method, the evaluation of the
VdW corrections to the interfragment forces is trivial, thus
allowing an easy implementation in standard geometry optimization
calculations or Molecular Dynamics simulations.
Remarkably, the whole procedure of generating the MLWFs and evaluating the
VdW corrections represents a negligible additional computational cost,
compared to that of a standard DFT calculation. 

We have applied the new method to selected dimers among typical VdW-bonded 
systems: Ar$_2$, N$_2$-N$_2$ (``T-shaped''),
CH$_4$-CH$_4$, C$_6$H$_6$-C$_6$H$_6$ (``sandwich-shaped''),
C$_6$H$_6$-Ar, CO$_2$-CO$_2$, and also
a mixed (H-bonded/VdW-bonded) complex, C$_6$H$_6$-H$_2$O.

In the Tables I-III we report our computed binding energy
(a positive value indicates unbound complexes), equilibrium characteristic 
interdimer distance, and fragment-fragment effective 
C$_6$ coefficient, which is defined as $\sum_{n,l} C_{6nl}$ 
(in fact, neglecting the differences in the spreads of the
MLWFs and in the interfragment distances between the WFCs,
$E_{\rm VdW} \sim -f\left( \sum_{n,l} C_{6nl} \right) /R^6$,
where $R$ is the interdimer distance).
These values are compared to the most
reliable (to our knowledge) reference literature corresponding data,
which are often spread over a relatively large range (comparison
with literature C$_6$ coefficients should be taken as purely
indicative because of different assumed definition).
As can be seen, the general performance of the method is quite
satisfactory; in fact, the improvement achieved by including the VdW
correction, with respect to the pure revPBE results, is
dramatic, even in the case of a mixed complex, such as C$_6$H$_6$-H$_2$O,
where some fraction of the binding energy is already given by
the standard DFT calculation.
In Fig. 1 we show the effect of the inclusion of the VdW correction
on the behavior of the binding energy of the Ar$_2$ dimer,
plotted as a function of the Ar-Ar distance, and compared
to the reference equilibrium value; in Ar the 8 valence
electrons of each atom are described by 4 MLWFs
(spin degeneracy is exploited), whose WFCs are tetrahedrally located around
the Ar ion. 

In the case of the equilibrium characteristic distances, the more
substantial deviation from the reference results can easily be
explained by the fact that the potential energy curves for
weakly-bonded systems are typically very shallow.  
Inspection of Table I shows that anisotropy effects do not
much affect the binding energy estimates, but for the case of
the ``sandwich-shaped'' benzene dimer, where anisotropy correction
leads to a value much closer to the reference data: this behavior
comes as no surprise, since the planar geometry of the two
fragments clearly induces strong anisotropy in the computed
MLWFs. 
In Table I, by considering the MLWF anisotropy, the binding energies
are always decreased and lower than the
reference values; this behavior is probably due to the neglect
of higher-order contributions to the VdW correction, such as
the $-C_8/R^8$ term (dipole-quadrupole interaction), which
should be included to have a very accurate estimate 
of the binding energy\cite{Adamovic}.

We have also applied our technique to the case of graphite,
where the optimal interlayer distance, found to be
8.69 \AA\ with the standard DFT-revPBE approach, becomes 6.33 \AA\
by including VdW effects, that is much closer to the
experimental value (6.70 \AA).

In conclusion, we have presented and applied in test cases
a technique suitable to describe VdW effects in the framework
of standard DFT calculations. The technique is based on the
generation of the MLWFs and naturally describes changes
in the electronic density distributions of the fragments
due to the environment, for instance related to 
charge polarization effects: in fact these changes are 
easily described in terms of changes in the location of
the centers and in the spreads of the MLWFs.
The results of the method,
which is simple to be implemented and not expensive computationally,
are quite satisfactory and promising, also considering that
a large area for future improvements exists: in fact, different, more
sophisticated schemes to utilize the MLWFs could be developed and/or
improved reference DFT functionals, with respect to revPBE, could
be adopted.   

We thank F. Ancilotto, M. Boero, F. Toigo, and F. Valencia for
useful discussions. We acknowledge funding from
Padua University project n. CPDA033545 and
PRIN 2004 (``NANOTRIBOLOGIA'' project),
and allocation of computer resources from INFM
``Progetto Calcolo Parallelo''.

\vfill
\eject

\begin{table}
\caption{Binding energy, in meV, computed using the standard DFT-revPBE
calculation, $E_0$ and including the VdW correction, $E_0 + E_{\rm VdW}$,
compared to available literature data; in parenthesis values, computed
taking anisotropy effects into account, are reported.}
\begin{center}
\begin{tabular}{|c|c|c|c|}
\hline
system & $E_0$ & $E_0 + E_{\rm VdW}$ & ref \\ \tableline
\hline
Ar-Ar                 &  -1.7 &  -11.9  (-9.5) &  -12.3 \\ 
N$_2$-N$_2$           &  -2.9 &  -11.1 (-10.8) &  -13.3 \\
CH$_4$-CH$_4$         &  -2.1 &  -11.7  (-9.9) &  -23$\leftrightarrow$-14 \\
C$_6$H$_6$-C$_6$H$_6$ &  +7.1 & -252.7 (-144.1)&  -142$\leftrightarrow$-61 \\
C$_6$H$_6$-Ar         &  -2.4 &  -65.7 (-53.5) &  -65 \\
CO$_2$-CO$_2$         & -16.2 &  -54.0 (-47.9) &  -69$\leftrightarrow$-59 \\ 
C$_6$H$_6$-H$_2$O     & -40.4 & -131.2 (-121.6)& -169$\leftrightarrow$-137 \\
\hline
\end{tabular}                                                
\end{center}
\label{table1}                                  
\end{table}

\begin{table}
\caption{Equilibrium characteristic interdimer distance, in \AA, 
computed using the standard DFT-revPBE
calculation, $R_0$ and including the VdW correction, $R_{\rm VdW}$,
compared to available literature data; in parenthesis values, computed
taking anisotropy effects into account, are reported.}
\begin{center}
\begin{tabular}{|c|c|c|c|}
\hline
system & $R_0$ & $R_{\rm VdW}$ & ref \\ \tableline
\hline
Ar-Ar                 &  4.67 &   4.03 (4.07) &  3.76 \\ 
N$_2$-N$_2$           &  5.05 &   4.37 (4.37) &  4.03 \\
CH$_4$-CH$_4$         &  4.70 &   4.23 (4.25) &  3.60$\leftrightarrow$4.27 \\
C$_6$H$_6$-C$_6$H$_6$ &   --- &   3.45 (3.45) &  3.80$\leftrightarrow$3.90 \\
C$_6$H$_6$-Ar         &  4.79 &   3.57 (3.57) &  3.41 \\
CO$_2$-CO$_2$         &  3.86 &   3.49 (3.49) &  3.60 \\ 
C$_6$H$_6$-H$_2$O     &  4.23 &   3.17 (3.17) &  3.40$\leftrightarrow$3.50 \\
\hline
\end{tabular}                                                
\end{center}
\label{table2}                                  
\end{table}

\begin{table}
\caption{Fragment-fragment effective $C_6$ coefficient (see text for the
definition), in a.u., computed 
using the standard DFT-revPBE with the VdW correction,
compared to available literature data; in parenthesis values, computed
taking anisotropy effects into account, are reported.}
\begin{center}
\begin{tabular}{|c|c|c|}
\hline
system & $C_6$ &  ref \\ \tableline
\hline
Ar-Ar                 &    92.5 (74.7)  &    64.3$\leftrightarrow$65.5 \\
N$_2$-N$_2$           &    90.3 (95.6)  &    73.4 \\
CH$_4$-CH$_4$         &   103.0 (101.0) &   118.0$\leftrightarrow$130.0 \\ 
C$_6$H$_6$-C$_6$H$_6$ &  2930.0 (2460.0)&  1723.0 \\
C$_6$H$_6$-Ar         &   490.0 (448.0) &   330.1 \\
CO$_2$-CO$_2$         &   187.0 (172.0) &    --- \\
C$_6$H$_6$-H$_2$O     &   323.0 (299.0) &   208.5$\leftrightarrow$277.4 \\
\hline
\end{tabular}                                                
\end{center}
\label{table3}                                  
\end{table}

\vfill
\eject

\pagestyle{empty}
\begin{figure}
{\vskip 1.3cm}
\centerline{
\includegraphics[width=17cm]{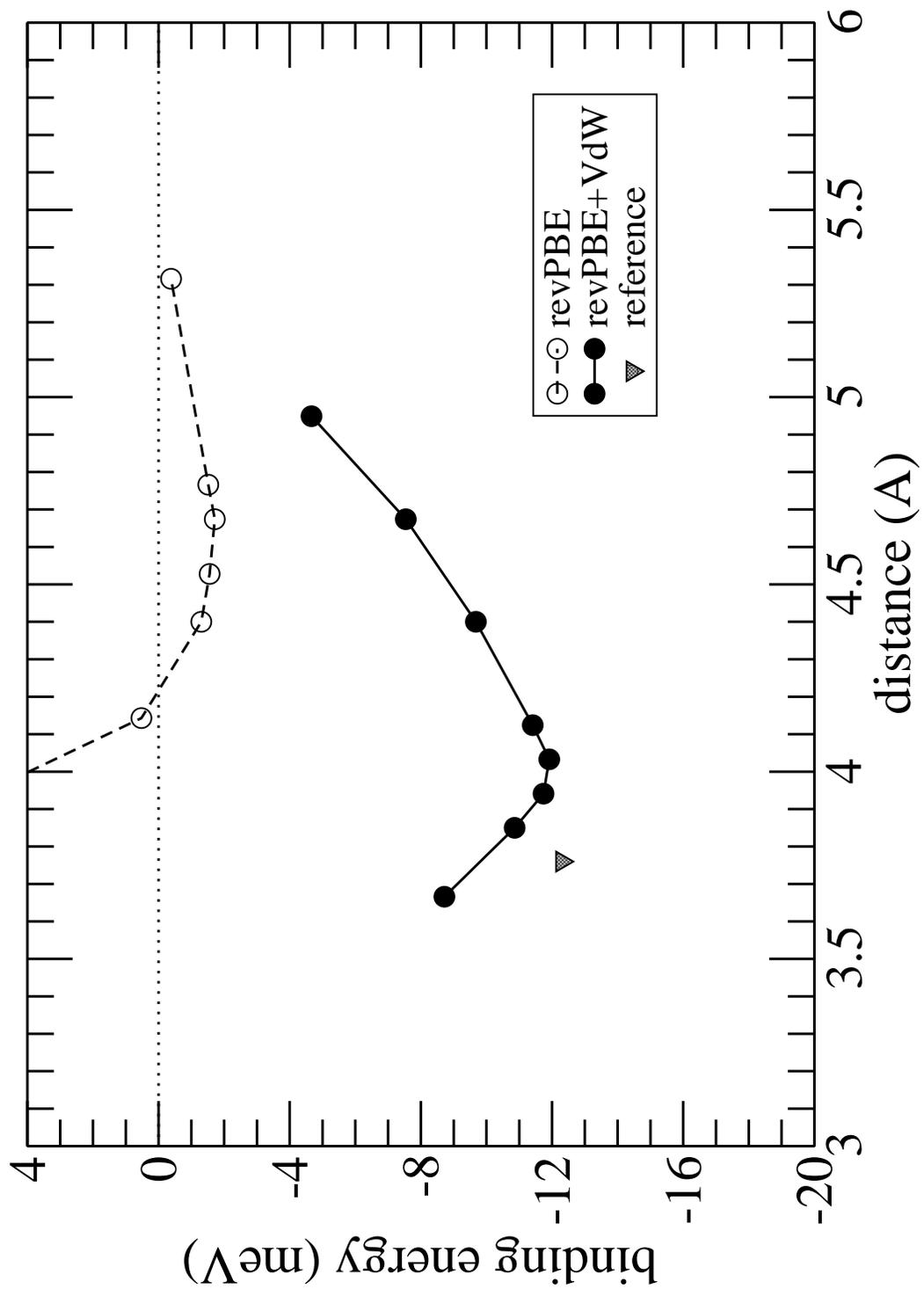}
}
\caption{Binding energy of the Ar$_2$ dimer, as a function of the 
Ar-Ar distance, using the standard DFT-revPBE calculation and 
including the VdW correction.}
\label{fig1}
\huge
\end{figure}

\end{document}